\documentclass[submission,copyright,creativecommons]{eptcs}
\providecommand{\event}{TTATT 2013} % Name of the event you are submitting to
\usepackage{breakurl}             % Not needed if you use pdflatex only.

\usepackage{times}
\usepackage[dvips]{graphicx}
\usepackage{amsmath}
\usepackage{amssymb}
\usepackage{enumerate}

\newtheorem{define}{Definition}
\newtheorem{example}{Example}

\newenvironment{proof}{{\it Proof.}\quad}{}

\makeatletter
\def\@begintheorem#1#2{\trivlist \item[\hskip \labelsep{\bf #1\ #2:}]}
\def\@opargbegintheorem#1#2#3{\trivlist
    \item[\hskip \labelsep{\bf #1\ #2\ (#3):}]}
\makeatother

\title{Toward Security Verification against Inference Attacks on Data Trees}
\author{Ryo Iwase \qquad Yasunori Ishihara \qquad Toru Fujiwara
\institute{Graduate School of Information Science and Technology, Osaka University}
\email{\{r-iwase,ishihara,fujiwara\}@ist.osaka-u.ac.jp}
}
\def\titlerunning{Toward Security Verification against Inference Attacks on Data Trees}
\def\authorrunning{Ryo Iwase, Yasunori Ishihara, \& Toru Fujiwara}
\begin{document}
\maketitle

\begin{abstract}
This paper describes our ongoing work on security verification
against inference attacks on data trees. 
We focus on infinite secrecy against inference attacks,
which means that attackers cannot narrow
down the candidates for the value of the sensitive information to
finite by available information to the attackers.
Our purpose is to propose a model under which infinite secrecy is
decidable.
To be specific,
we first propose tree transducers which are expressive enough to represent
practical queries.
Then, in order to represent attackers' knowledge,
we propose data tree types 
such that type inference and inverse type inference on those tree
transducers are possible with respect to data tree types,
and infiniteness of data tree types is decidable.
\end{abstract}

%introduction
\section{Introduction}
Nowadays, many organizations utilize and store information in databases.
These databases may contain highly confidential information.
One of the important problems on achieving database security for these database systems
is to ensure the security against inference attacks.
Inference attacks mean that users infer the information which they cannot access directly
by using the authorized queries and the result of them. 
In order to ensure the security of databases,
it is important to figure out the possibility in advance that the sensitive information can be leaked by inference attacks.
\begin{example}
%\rm{}
We show an example of inference attacks on XML databases.
We consider an XML document $I$ (see Fig. \ref{fig:instance}) representing the correspondence between student name,
origin, and the amount of the scholarship, and valid against the following schema:
\begin{eqnarray*}
{\sf faculty} & \rightarrow & {\sf (student\{@scholarship\})}^*
\\
{\sf student\{@scholarship\}} & \rightarrow & {\sf name\{@str\}, origin\{@str\}}
\end{eqnarray*}
That is, the {\sf faculty} element has zero or more {\sf student} elements as its children.
Each {\sf student} element has a {\sf scholarship} as a data value which is a nonnegative integer without upper limit, and
has a {\sf name} element and an {\sf origin} element as its children.
Each of {\sf name} and {\sf origin} element has a {\sf str} as a string data value.

Let $T_1$ be an authorized query extracting the name and the origin of each student.
Let $T_2$ be an authorized query extracting the origin of the student who receives
the most amount of the scholarship, and $T_3$ be an authorized query extracting the origin and scholarship
of the student who receives the second most amount of the scholarship.
Moreover, we set the sensitive information to the amount of the scholarship which a student of a given name receives, and
let $T_{\rm A}$, $T_{\rm B}$, $T_{\rm C}$, and $T_{\rm D}$ be the unauthorized queries extracting
the amount of the scholarship of 
the student of the name A, B, C, and D, respectively (i.e., extracting the sensitive information).
Now, we assume that the results of $T_1$, $T_2$, and $T_3$ are the trees shown in Figs. \ref{fig:name_born},
\ref{fig:maxscholar}, and \ref{fig:2ndscholar}, respectively.
%\begin{figure}[htbp]
%\centering
%\rotatebox[origin=c]{0}{
%\includegraphics[scale=.25]{figure/name_born.eps}}
%\caption{The result of $T_1$.}
%\label{fig:name_born}
%\end{figure}

\begin{figure}[htbp]
\centering
\rotatebox[origin=c]{0}{
\includegraphics[scale=.6]{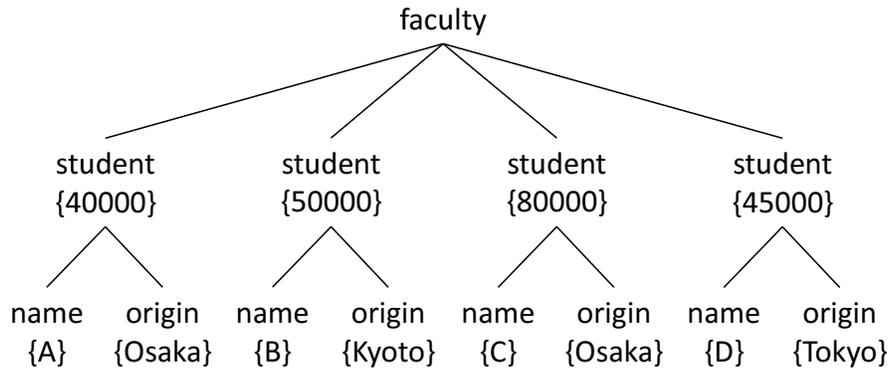}}
\caption{An XML Document $I$.}
\label{fig:instance}
\end{figure}
\begin{figure}[htbp]
\centering
\rotatebox[origin=c]{0}{
\includegraphics[scale=.6]{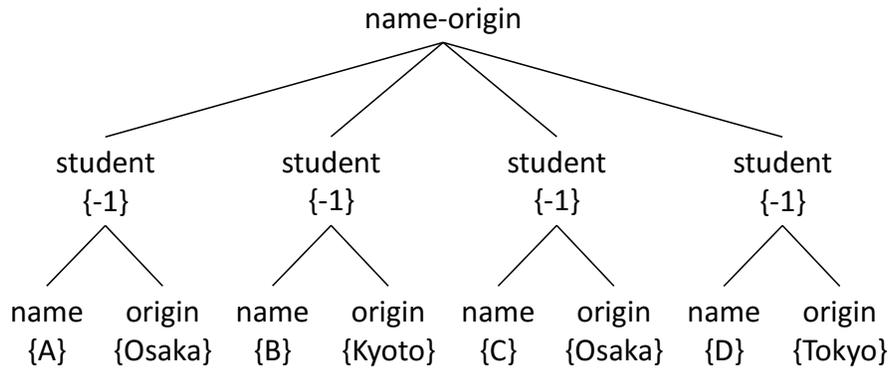}}
\caption{The result of $T_1$.}
\label{fig:name_born}
\end{figure}
\begin{figure}[htbp]
\begin{tabular}{cc}
\begin{minipage}{0.5\hsize}
\begin{center}
\rotatebox[origin=c]{0}{
\includegraphics[scale=.6]{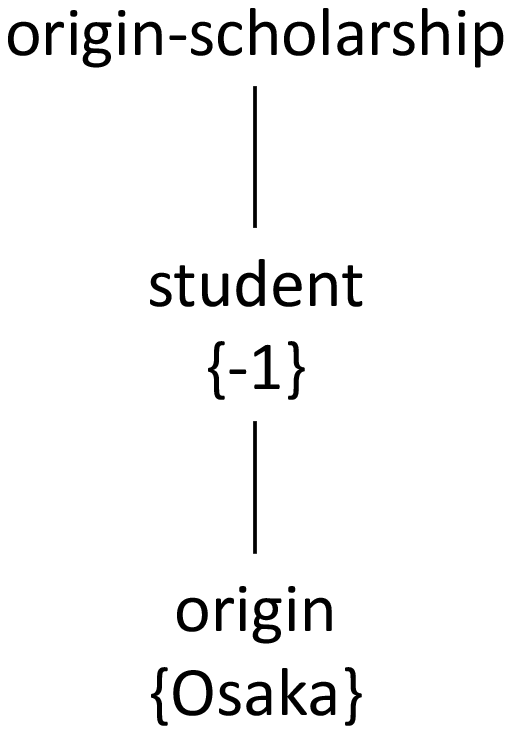}}
\caption{The result of $T_2$.}
\label{fig:maxscholar}
\end{center}
\end{minipage}
\begin{minipage}{0.5\hsize}
\begin{center}
\rotatebox[origin=c]{0}{
\includegraphics[scale=.6]{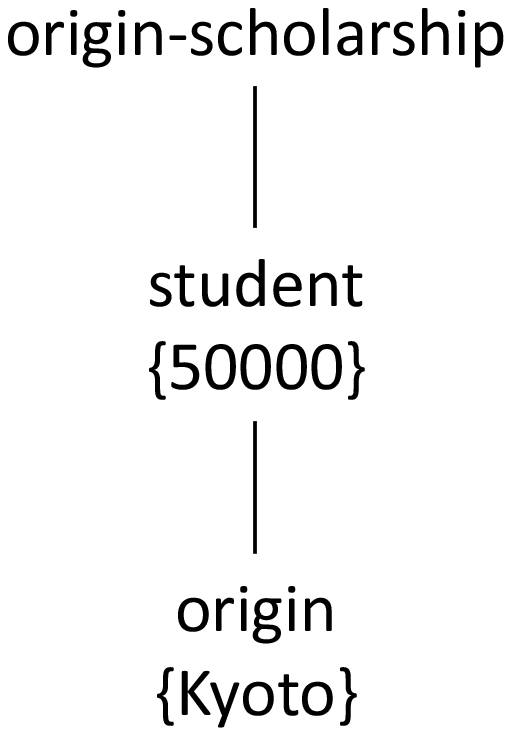}}
\caption{The result of $T_3$.}
\label{fig:2ndscholar}
\end{center}
\end{minipage}
\end{tabular}
\end{figure}
%\begin{figure}[htbp]
%\centering
%\rotatebox[origin=c]{0}{
%\includegraphics[scale=.25]{figure/sensitive.eps}}
%\caption{Identified sensitive information.}
%\label{fig:sensitive}
%\end{figure}

%\begin{figure}[htbp]
%\begin{tabular}{cc}
%\begin{minipage}{0.43\hsize}
%\begin{center}
%\rotatebox[origin=c]{0}{
%\includegraphics[scale=.32]{figure/sensitive_eng.eps}}
%\caption{Identified sensitive information.}
%\label{fig:sensitive}
%\end{center}
%\end{minipage}
%\begin{minipage}{0.43\hsize}
%\begin{center}
%\rotatebox[origin=c]{0}{
%\includegraphics[scale=.32]{figure/born_2ndscholar_eng.eps}}
%\caption{The result of $T_3$.}
%\label{fig:2ndscholar}
%\end{center}
%\end{minipage}
%\begin{minipage}{0.53\hsize}
%\begin{center}
%\rotatebox[origin=c]{0}{
%\includegraphics[scale=.32]{figure/pattern1and2.eps}}
%\caption{Two candidates of the sensitive information.}
%\label{fig:pattern1and2}
%\end{center}
%\end{minipage}
%\end{tabular}
%\end{figure}
Then, we know that from the result of $T_1$, the student whose origin is Kyoto is only B, and
from the result of $T_3$, the student whose origin is Kyoto receives 50,000 yen as a scholarship.
Therefore, we find that B receives 50,000 yen.
That is, the result $T_{\rm B}(I)$ of $T_{\rm B}$ is identified by inference attacks.
Moreover, from the result of $T_2$, we know that a student whose origin is Osaka receives the most amount of the scholarship.
Here, considering that the origin of the student who receives the most amount of the scholarship is Osaka,
we find that D receives less than 50,000 yen.
Therefore, we can narrow down the number of the candidates of $T_{\rm D}(I)$ to 50000.
However, we cannot identify the person who receives the most amount of the scholarship
because we know that there are two students whose origins are Osaka.
Also, we do not know the most amount of the scholarship.
Therefore, we cannot narrow down the number of the candidates of
each of $T_{\rm A}(I)$ and $T_{\rm C}(I)$ to a finite number. \mbox{}\hfill $\square$
\end{example}
%
%Recently, inference attacks on relational databases has been studied in the context of privacy protection.
%Ref.~\cite{Sw02} proposes the security called $k$-anonymity against ``linking attacks''.
%$k$-anonymity means that for each tuple of a table, focused on quasi-identifier,
%there exists at least $k-1$ tuples which cannot be differentiated from the tuple.
%Ref.~\cite{MGKV06} proposes the stronger security definition called $l$-diversity.
%$l$-diversity means that a table satisfies $k$-anonymity and there exists at least $l$ candidates of
%the value of the sensitive information inferred by attacks.

The protection of sensitive information in XML databases has been studied in terms of access control.
In~\cite{DVPS02}, an access control model to protect information is proposed.
In the model, information to be protected is represented by a path expression, and for each information,
the authorizations of users are defined clearly.
Ref.~\cite{Ch13} discusses access control in the presence of insertions and updates of a database. 

\begin{figure}[htbp]
\centering
\rotatebox[origin=c]{0}{
\includegraphics[scale=.6]{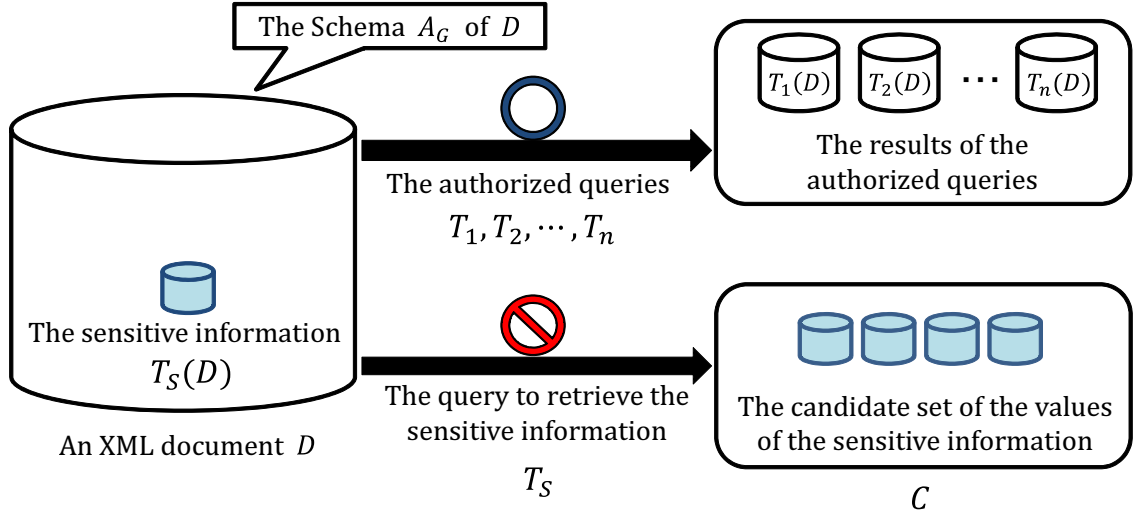}}
\caption{The whole picture of our verification.}
\label{fig:verification}
\end{figure}

In our previous work~\cite{HSTIF09}, we formulated the security against inference attacks on
XML databases and proposed a verification method of the security called infinite secrecy.
The whole picture of our verification is shown in Fig. \ref{fig:verification}.
The notion of infinite secrecy is as follows:
Suppose that the following information is available to a user:
\begin{itemize}
\item The authorized queries $T_1$, $T_2$, $\ldots$, $T_n$,
\item The results $T_{1}(D)$, $T_{2}(D)$, $\ldots$, $T_{n}(D)$ of the authorized queries on an XML document $D$,
\item The schema $A_G$ of $D$, and
\item The query $T_S$ to retrieve the sensitive information.
\end{itemize}
Then, the candidate set $C$ of the values of the sensitive information inferred by the user is
\begin{eqnarray*}
C & = & \{ T_{S}(D^\prime ) \mid D^\prime \in TL(A_{G}), T_{1}(D^\prime ) = T_{1}(D),
T_{2}(D^\prime ) = T_{2}(D), \ldots , T_{n}(D^\prime ) = T_{n}(D) \},
\end{eqnarray*}
where $TL(A_{G})$ denotes the set of trees valid against $A_{G}$.
If $|C|$ is infinite, then we say that \emph{$D$ is infinitely secret with respect to $T_S$}.
In the example above, $I$ is infinitely secret with respect to $T_{\rm A}$ and $T_{\rm C}$, and is not
with respect to $T_{\rm B}$ and $T_{\rm D}$.
The proposed verification method can only handle queries represented by relabeling or deleting the specified nodes
in an XML document. Since the set of labels is finite in the formulation,
the verification method of the security with queries involving infinite data value comparisons has not been studied.

We consider the verification of infinite secrecy against inference attacks on data trees.
The verification method consists of the following three steps:
\begin{enumerate}
\item Construct a candidate set of XML document $D$ from the authorized queries $T_{1}$, $T_{2}$, $\ldots$,
$T_{n}$, their results $T_{1}(D)$, $T_{2}(D)$, $\ldots$, $T_{n}(D)$, and the schema $A_{G}$.
\item Construct a candidate set $C$ of the value of $T_{S}(D)$ from the candidate set of $D$ and $T_{S}$.
\item Decide whether the number of the elements of $C$ is infinite.
\end{enumerate}
In the verification, \emph{type inference} and \emph{inverse type inference} on tree transducers are used at steps 2 and 1,
respectively.
\emph{Type inference} is to construct the candidate set $Doc^{\prime}_{out}$ of output trees of a tree transducer
from the tree transducer and the candidate set $Doc_{in}$ of input trees.
\emph{Inverse type inference} is to construct the candidate set $Doc^{\prime}_{in}$ of input trees of a tree transducer
from the tree transducer and the candidate set $Doc_{out}$ of output trees.
To verify the security according to these steps, the models must satisfy three requirements.
First, tree transducers must be able to represent practically significant queries.
Second, type inference and inverse type inference on tree transducers must be possible.
Third, it must be decidable whether the number of the candidates of the sensitive information is infinite.

In this paper, we discuss a verification method of the security 
against inference attacks on data trees.
We propose models satisfying the aforementioned three requirements.
First, we propose tree transducers on data trees which are expressive enough to represent practical queries.
Operations corresponding to projection, selection, and natural join
in the relational algebra are allowed in queries by using those tree transducers.
Then, in order to represent attackers' knowledge, we propose \emph{data tree types}
such that type inference and inverse type inference on those tree transducers
are possible with respect to data tree types, and infiniteness of data tree types is decidable.
A data tree type consists of a non-deterministic finite tree automaton, 
a mapping from the set of the pairs of the states of the tree automaton and the labels of the nodes to the set of variables,
and a finite set of conditional expressions between variables or between a variable and a constant.
Until now, we have provided inverse type inference on several tree transducers
and type inference on data-rewriting transducers, and have provided an algorithm to decide infiniteness of data tree types.
%However, the details of type inference, inverse type inference and the decision of infiniteness of data tree types
%are omitted in this paper because of space limitation.

%existing models
\section{Preliminaries}
We use \emph{data trees}~\cite{MATL09}\cite{NTFDV01} as a model of XML documents.
We define data trees as follows.
\begin{define}
Let $\Sigma$ be a finite set of labels including special symbols $\sharp$ and $\$$,
and $D$ be a countable infinite set of data values
on which a total order $<$ is defined.
We assume that $D$ is a set of integers or rational numbers.
A data tree $t$ is a 3-tuple $\langle T, l, \rho \rangle$, where
\begin{itemize}
\item $T$ is a set of nodes, which is a prefix-closed finite subset of $\mathbb{N}^{*}$
such that for all $j<i$ and $v\in \mathbb{N}^{*}$, if $v \cdot i\in T$ then $v \cdot j\in T$,
\item $l$ is a mapping from $T$ to $\Sigma$, and
\item $\rho $ is a mapping from $T$ to $D$.
\end{itemize}
That is, each node has just one label and one data value. 
In $T$, $\epsilon$ is called the root node, and for any two nodes $v$, $v \cdot i\in T$,
$v$ is called the parent of $v \cdot i$, and $v \cdot i$ is called the $i$-th child of $v$.
\mbox{}\hfill $\square$
\end{define}
We use \emph{non-deterministic finite tree automata} (NFTAs) to represent XML schemas.
%First, we define a non-deterministic finite automaton (NFA) used in the definition of NFTAs.
%\begin{define}
%An NFA $e$ is a 5-tuple $(P, N, \delta, \hat{p}, P_f)$, where
%\begin{itemize}
%\item $P$ is a finite set of states,
%\item $N$ is an alphabet,
%\item $\delta$ is a set of transition rules in the form of $(p, a, p^\prime)$ where $p,p^\prime \in P$ and $a \in N$,
%\item $\hat{p} \in P$ is the initial state, and
%\item $P_f \subseteq P$ is a set of final states.
%\end{itemize}
%Given an NFA $e$ over $N$, let $L(e)$ denote the language over $N$ recognized by $e$. \mbox{}\hfill $\square$
%\end{define}
%Now, we define an NFTA as follows.
\begin{define}
An NFTA $A$ is a 4-tuple $(Q, \Sigma , q_{0}, R)$, where
\begin{itemize}
\item $Q$ is a finite set of states,
\item $\Sigma $ is a finite set of labels,
\item $q_{0} \in Q$ is the initial state, and
\item $R$ is a set of transition rules in the form of $(q, a, e)$, where $q \in Q$, $a \in \Sigma$, and $e$ is a
non-deterministic finite automaton over $Q$.
\end{itemize}
A run of an NFTA assigns states to nodes of an input tree according to the transition rules.
Formally, we define a run $r^t_A$ of $A=(Q, \Sigma , q_0 , R)$ against $t=\langle T, l, \rho \rangle$ 
as a mapping from $T$ to $Q$ with the following properties:
\begin{itemize}
\item $r^t_A(\epsilon )= q_0$.
\item For each node $v\in T$ with $n$ children, there exists a transition rule $(q,a,e)\in R$ such that
$r^{t}_{A}(v)=q$, $l(v)=a$, and $r^{t}_{A}(v \cdot 1)r^{t}_{A}(v \cdot 2)\cdots r^{t}_{A}(v \cdot n)$ is in the string language
represented by $e$.
\end{itemize}
$t$ is accepted by $A$ if there exists a run of $A$ against $t$.
Let $TL(A)$ denote the set of data trees accepted by $A$.
\mbox{}\hfill $\square$\\
\end{define}%document model, schema model

%proposed models
\section{Proposed Models}
%In this section, we define the models of XML databases for the verification of the security.
\subsection{Queries}
We use deterministic tree transducers to represent queries.
We define seven types of tree transducers for our verification.

\begin{itemize}
\item A \emph{deterministic top-down relabeling tree transducer}~\cite{HSTIF09} relabels the current node
according to the state of the node and assigns states to the children of the node, traversing in a top-down manner. %top-down relabeling tree transducer
\item A \emph{deterministic bottom-up relabeling tree transducer}~\cite{HSTIF09} relabels the current node
according to the label of the node and the states of the children, traversing in a bottom-up manner.%bottom-up relabeling tree transducer
\item A \emph{deterministic deleting tree transducer}~\cite{HSTIF09} deletes nodes labeled by $\sharp$ and subtrees
rooted by $\$$. %deleting tree transducer

\item A \emph{deterministic data-rewriting tree transducer} rewrites the data value of all the nodes
which have a specified label $a$ to a specified value $d$.
The operation by a data-rewriting tree transducer corresponds to projection in the relational algebra.
Formally, given label $a$ and data value $d$, a deterministic data-rewriting tree transducer
transforms $t = \langle T,l,\rho \rangle$ into $t^{\prime} = \langle T,l,\rho^\prime \rangle$,
where $\rho^\prime$ is a mapping defined as follows:
\begin{eqnarray}
\rho^{\prime}(v) & = & \left \{
\begin{array}{l@{\,\,\,}l}
d & \mbox{if }l(v) = a, \\
\rho (v) & \mbox{otherwise.}
\end{array}
\right. \nonumber
\end{eqnarray}

\item A \emph{deterministic data-relabeling tree transducer} relabels all the nodes which have a specified label $a$
and data value $d$ to a specified label $a^\prime$.
The operation by a data-relabeling tree transducer corresponds to selection in the relational algebra.
Formally, given labels $a$, $a^\prime$, and data value $d$, a deterministic data-relabeling tree transducer
transforms $t = \langle T,l,\rho \rangle$ into $t^\prime=\langle T,l^\prime,\rho  \rangle$,
where $l^\prime$ is a mapping defined as follows:
\begin{eqnarray}
l^{\prime }(v) & = & \left \{
\begin{array}{l@{\,\,\,}l}
a^\prime & \mbox{if } l(v) = a \mbox{ and } \rho (v) = d, \\
l(v) & \mbox{otherwise.}
\end{array}
\right. \nonumber
\end{eqnarray}

\item A \emph{deterministic min-data-relabeling tree transducer} relabels all the nodes which have the minimum value
of the nodes labeled by a specified label $a$ to a specified label $a^\prime$.
A deterministic min-data-relabeling tree transducer is used for representing an operation like natural join
in the relational algebra.
Formally, given labels $a$ and $a^\prime$, a deterministic min-data-relabeling tree transducer
transforms $t=\langle T,l,\rho \rangle$ into $t^\prime=\langle T,l^\prime ,\rho \rangle$, 
where $l^\prime$ is a mapping defined as follows:
\begin{eqnarray}
l^{\prime}(v) & = & \left \{
\begin{array}{l@{\,\,\,}l}
a^\prime & \mbox{if } l(v) = a \mbox{ and } \rho (v) = \mbox{min} \{ \rho (v^\prime ) 
\mid l(v^\prime )=a \} , \\
l(v) & \mbox{otherwise.}
\end{array}
\right. \nonumber
\end{eqnarray}

\item A \emph{deterministic max-data-relabeling tree transducer} is a counterpart of
a deterministic min-data-relabeling tree transducer.
Formally, given labels $a$ and $a^\prime$, a deterministic max-data-relabeling tree transducer
transforms $t=\langle T,l,\rho \rangle$ into $t^\prime=\langle T,l^\prime ,\rho \rangle$, 
where $l^\prime$ is a mapping defined as follows:
\begin{eqnarray}
l^{\prime}(v) & = & \left \{
\begin{array}{l@{\,\,\,}l}
a^\prime & \mbox{if }l(v) = a \mbox{ and } \rho (v) = \mbox{max}\{ \rho (v^\prime ) 
\mid l(v^\prime ) = a \} , \\
l(v) & \mbox{otherwise.}
\end{array}
\right. \nonumber
\end{eqnarray}

\end{itemize}%proposed tree transducers
%\subsubsection{Query Model for Verification}
The procedures of natural join by min/max-data-relabeling tree transducers are as follows.
We consider an XML document which have the information of two relations $A$ and $B$.
First, choose a pair of nodes $p$ and $q$, where $p$ and $q$ correspond to a tuple of $A$ and $B$, respectively.
Second, relabel their nodes to a new label $a^\prime$ by a bottom-up/top-down relabeling tree transducer.
Third, relabel all the nodes labeled by $a^\prime$ to a new label $b^\prime$ by a min/max-data-relabeling tree transducer.
If the values of $p$ and $q$ are the same, then both $p$ and $q$ are labeled by $b^\prime$, and we can join these two nodes.
Otherwise, we need to choose another pair because we cannot join these two nodes.

A {\it query} is a composition of these tree transducers
satisfying the following restrictions.
%First, the label to which either a data-relabeling tree transducer, a min-data-relabeling tree transducer, or
%a max-data-relabeling tree transducer relabels some label is not appeared in the input tree of the tree transducer.
First, every query must be a composition of
zero or more tree transducers except deleting tree transducers followed by a deleting tree transducer.
%%% Second, the unauthorized query $T_S$ must be compositions of zero or more
%%% relabeling tree transducers and data-rewriting tree transducers followed by a deleting tree transducer.
Second, no constituent tree transducers of
the unauthorized query $T_S$ relabels a node to $\sharp$.
These restrictions are necessary for type inference to be possible.
Without these restrictions, candidates for the value of the sensitive information cannot be represented
by an NFTA even if we do not consider data values.
For example, consider an NFTA $A = (Q, \Sigma, q_0, R)$, where
\begin{itemize}
\item $Q = \{ q_0, q_a, q_b \}$, and
\item $R = \{ (q_0, r, q_a q_b q_a), (q_a, a, \epsilon), (q_b, b, q_a q_b q_a | \epsilon) \}$.
\end{itemize}
For trees accepted by $A$, consider the trees obtained by relabeling the nodes labeled by $b$ to $\sharp$
and then deleting the nodes labeled by $\sharp$.
The resulting trees have the root node labeled by $r$, and the strings obtained by the concatenation of labels of its children is the form of $a^n b a^n$,
which cannot be represented by an NFTA.

Queries appeared in Example 1 can be represented by the proposed tree transducers.
For example, $T_{\rm A}$ can be represented as follows.
\begin{enumerate}
\item Relabel the {\sf name} nodes which have the value ``A'' to {\sf name}$^\prime$ by a data-relabeling tree transducer.
\item Relabel the {\sf student} nodes which have a {\sf name} node as a child to $\$$ by a bottom-up relabeling tree transducer.
\item Relabel the {\sf name}$^\prime$ nodes and {\sf origin} nodes to $\sharp$ by a top-down/bottom-up relabeling tree tranducer.
\item Delete nodes by a deleting tree transducer.
\end{enumerate}
The others can also be represented similarly.
%\begin{figure}[htbp]
%\begin{minipage}{\hsize}
%\centering
%\rotatebox[origin=c]{0}{
%\includegraphics[scale=0.25]{figure/querymodel.eps}}
%\caption{a}(B}
%\label{fig:query}
%\end{minipage}
%\end{figure}%query model for verification
\subsection{Data tree types}

We introduce \emph{data tree types} to represent sets of data trees,
which model user's knowledge during inference attacks.
A data tree type is defined as a finite union of
\emph{atomic data tree types}.
Similarly to existing research on incomplete information like~\cite{ASV06},
we use variables and conditional expressions on them
to represent undetermined data values.
However, there are several novelties in our model.
First, each variable is associated with a pair of
a state of an NFTA and a label,
rather than a node of a fixed tree.
Hence, one atomic data tree type can handle infinitely many variations of
tree shapes (up to the expressive power of NFTAs).
This is useful for type inference involving relabeling and
deleting tree transducers.
Next, our model uses two kinds of variables.
\emph{S-variables} are ordinary ones, and
all data values of nodes which have the same s-variable
must be the same.
\emph{M-variables} are novel ones, and
all data values of nodes which have the same m-variable are not necessarily
the same, but satisfy conditional expressions on the m-variable.
M-variables are useful for inverse type inference involving
data-rewriting tree transducers.
For example, consider inverse type inference on data-rewriting tree transducers
which rewrites the data value of all the nodes labeled by $a$ to value $d$.
We cannot know data values of the nodes labeled by $a$ in input trees since
these values have already been rewritten to $d$.
By the following definition of $\theta$, these nodes may have the same variable
if a run of an NFTA assigns the same state to these nodes.
On the other hand, these nodes may have different values.
Since m-variable can have more than one value,
we can represent values of these nodes by an m-variable. 

In what follows, s-variables and m-variables are written as
$\ddot{x}$ and $\tilde{x}$, respectively.
We also write simply like $x$
when we do not differentiate two kinds of variables.

\begin{define}
An \emph{atomic data tree type} $Doc$
is a 3-tuple $\langle A, \theta , E\rangle$,
where
\begin{itemize}
\item $A=(Q, \Sigma , q_{0}, R)$ is an NFTA;
\item $\theta$ is a mapping from $Q \times \Sigma$ to the set of
  variables; and
\item $E$ is a finite set of conditional expressions in the following form:
 \begin{itemize}
 \item $x$ op $\ddot{y}$, where op $\in \{ <,>,\subseteq \}$,
 \item $d$ op $y$, where op $\in \{ \in, \not\in \}$ and $d$ is a constant, and
% \item $\tilde{x}$ op $\ddot{y}$, where op $\in \{ <,>,\subseteq \}$,
% \item $\ddot{x}$ op $d$, where op $\in \{ =,\neq \}$ and $d$ is a constant, and
 \item $\tilde{x} \subseteq \tilde{y}$.
\mbox{}\hfill $\square$
 \end{itemize}
\end{itemize}
\end{define}

The semantics of an atomic data tree type is defined as follows.
Let $\sigma$ be a mapping from a set of variables to the power set of
values satisfying
the following conditions:
\begin{itemize}
\item for all s-variables $\ddot{x}$, $|\sigma (\ddot{x})| = 1$, and
\item for all m-variables $\tilde{x}$, $|\sigma (\tilde{x})| \geq 0$.
\end{itemize}
We extend the definition of $\sigma$ toward conditional
expressions as follows:
\begin{itemize}
\item $\sigma (x$ op $\ddot{y})=\{ a$ op $b \mid a\in \sigma (x), b\in \sigma (y) \}$, where op $\in \{ <, > \}$,
\item $\sigma (x \subseteq \ddot{y})=\{ \sigma(x) \subseteq \sigma(\ddot{y}) \}$,
\item $\sigma (d$ op $y) = \{ d$ op $\sigma (y) \}$, where op $\in \{ \in, \not\in \}$ and $d$ is a constant,
\item $\sigma (\tilde{x} \subseteq \tilde{y}) = \{ \sigma(\tilde{x}) \subseteq \sigma(\tilde{y}) \}$, and
\item $\sigma (E)=\displaystyle \bigcup_{x \mbox{ \small{op} }y \hspace{0.1pc}\in E}\sigma (x\mbox{ op }y)$.
\end{itemize}
We assume that $D$ is a set of integers or rational numbers.
Therefore, each meaning of $<$ and $>$ is the same as that in sets of integers or rational numbers.

A data tree $t=\langle T, l, \rho \rangle$ \emph{belongs to}
the set of data trees
represented by an atomic data tree type
$Doc=\langle A, \theta , E\rangle$
if there exist
a run $r^t_A$ and
a mapping $\sigma$ such that
for all $v\in T$, $\rho (v)\in \sigma ( \theta(r^{t}_{A}(v), l(v)) )$,
and
all the conditional expressions in $\sigma (E)$ hold.
Let $TL(Doc)$ denote the set of data trees which belongs to $Doc$.

%%% For $Doc=\langle A, \theta , E\rangle$, let $TL(Doc)$ denote the set of data trees in $Doc$. 
%%% Moreover, let $E[x\rightarrow x^\prime]$ denote the finite set of relational expressions which is obtained by 
%%% replacing each $x$ in $E$ with $x^\prime$. 

%%% The expression $\ddot{x}<\tilde{x}$ means that all data values of nodes which have variable $\tilde{x}$ are
%%% larger than the value of $\ddot{x}$.

%%% Local Variables:
%%% mode: LaTeX
%%% TeX-master: "main_eng.tex"
%%% TeX-command-default: "LaTeX"
%%% End:
%data tree types

%decidability
\section{Decidability}
In this section, we refer to the detail of type inference and inverse type inference,
and provide an algorithm to decide infiniteness of data tree types.
\subsection{Type inference and inverse type inference}
As stated already, we have proved the correctness of inverse type inference on several tree transducers
and type inference on data-rewriting transducers.
In this section, we show the detail of inverse type inference on data-rewriting tree transducers.

For a data-rewritng tree transducer which rewrites the data value of all the nodes
labeled by $a$ to value $d$,
the data tree type $Doc^\prime = \langle A^\prime, \theta^\prime, E^\prime \rangle $ of input trees is constructed from
the data tree type $Doc = \langle A, \theta, E \rangle $ of output trees, where $A = (Q, \Sigma, q_0, R)$.
The detail of $\langle A^\prime, \theta^\prime. E^\prime \rangle$ is as follows.
\begin{itemize}
\item $A^\prime = A$.
\item $\theta^\prime(q,c) = \left \{
\begin{array}{ll}
\tilde{x}^\prime & \mbox{if } c = a, \\
\theta(q,c) & \mbox{otherwise.}
\end{array}
\right. \nonumber$\\
Here, $\tilde{x}^\prime$ is a variable which do not appear in $Doc$.
\item $E^\prime = E \cup \{d \in \theta(q,a) \mid (q,a,e)\in R \}$.
\end{itemize}

%data-rewriting proof beginning
%%%%%%%%%%%%%%%%%%%%%%%%%%%%%%%%%%%%%%%%
\begin{proof}
First, let $t^\prime = \langle T,l,\rho^\prime \rangle \in TL(\langle A^\prime, \theta^\prime, E^\prime \rangle)$
be an input tree.
There exist a run $r^{t^\prime}_{A^\prime}$ and a mapping $\sigma^\prime$, and the following properties hold:
\begin{itemize}
\item for all $v \in T$, $\rho(v) \in \sigma^\prime(\theta^\prime(r^{t^\prime}_{A^\prime}(v), l(v)))$, and
\item all the conditional expressions in $\sigma^\prime (E^\prime)$ hold.
\end{itemize}
Then, by the definition of the tree transducer,
the output tree $t = \langle T, l, \rho \rangle$ is accepted by $A$
since there exists a run $r^t_A$ such that $r^t_A = r^{t^\prime}_{A^\prime}$, and
for all $v \in T$ satisfying $l(v) = a$, we have $\rho(v) = d$.
Here, let $\sigma = \sigma^\prime$.
Since all conditional expressions in $\sigma^\prime(E^\prime)$ hold,
for all $v \in T$ satisfying $l(v)=a$, $\rho(v) \in \sigma(\theta(r^t_A(v), l(v)))$.
Moreover, $E \subseteq E^\prime$.
Therefore, for $r^t_A$ and $\sigma$, the following properties hold:
\begin{itemize}
\item for all $v \in T$, $\rho(v) \in \sigma(\theta(r^t_A(v), l(v)))$, and
\item all the conditional expressions in $\sigma (E)$ hold.
\end{itemize}
Hence, $t \in TL(\langle A, \theta, E \rangle )$.

Inversely, let $t = \langle T,l,\rho \rangle \in TL(\langle A, \theta, E \rangle)$
be an output tree.
Then, for all $v \in T$ satisfying $l(v) = a$, $t$ must satisfy $\rho^\prime(v) = d$.
There exist a run $r^t_A$ and a mapping $\sigma$, and the following properties hold:
\begin{itemize}
\item for all $v \in T$, $\rho(v) \in \sigma(\theta(r^t_A(v), l(v)))$, and
\item all the conditional expressions in $\sigma (E)$ hold.
\end{itemize}
Then, by the definition of the tree transducer,
the input tree $t^\prime = \langle T,l,\rho^\prime \rangle$ is accepted by $A^\prime$
since there exists a run $r^{t^\prime}_{A^\prime}$ such that $r^{t^\prime}_{A^\prime} = r^t_A$.
Moreover, we define $\sigma^\prime$ as follows:
\begin{eqnarray*}
\sigma^\prime(x) = \left \{
\begin{array}{ll}
\{ \rho(v) \mid l(v) = a \} & \mbox{if } x = \tilde{x}, \\
\sigma(x) & \mbox{otherwise.}
\end{array}
\right. \nonumber
\end{eqnarray*}
Then, for all $v \in T$, $\rho(v) \in \sigma^\prime(\theta^\prime(r^{t^\prime}_{A^\prime}(v), l(v)))$.
Moreover, since for all $v \in T$ satisfying $l(v) = a$, $d \in \sigma(\theta(r^t_A(v), l(v)))$,
all conditional expression in $\sigma^\prime(E^\prime)$ hold.
Hence, $t^\prime \in TL(\langle A^\prime, \theta^\prime, E^\prime \rangle )$. \mbox{}\hfill $\square$
\end{proof}
%%%%%%%%%%%%%%%%%%%%%%%%%%%%%%%%%%%%%%%%
%data-rewriting proof end%decidability of type inference and inverse type inference
\subsection{Infiniteness of data tree types}
The following algorithm is to decide infiniteness of an input data tree type
$Doc = (A, \theta , E)$, where $A = (Q, \Sigma , q_{0}, R)$.
Before discussing the detail of the algorithm,
we declare in advance that we can assume that
there is no conditional expression which cannot be satisfied in $E$ for the following reason.
If there existed some conditional expressions which cannot be satisfied in $E$,
then by the definition of data tree types, $TL(Doc)$ would be an emptyset.
However, this assumption is contradictory since $TL(Doc)$ must contain at least the sensitive information.

The detail of the algorithm for deciding infiniteness is as follows.
\begin{enumerate}
\item Separate the number line into zones by constants in $E$.
If $D$ is an integer set, then we consider only integers in each zone.
\item Find the assignment $\sigma$ satisfying all conditional expressions in $E$ as follows.
\begin{enumerate}[2-1]
\item For each s-variable in $E$, assign a zone non-deterministically and then break up zones according to the assignment.
%For different s-variables to which the same zone is assigned, decide which of them is larger non-deterministically,
%and break up zones according to the relation.
\item Assign a set of zones to each m-variable in $E$ non-deterministically.
\end{enumerate}
%\item Check whether the assignment $\sigma$ at steps 2 and 3 satisfies all conditional expressions in $E$.
%If $\sigma$ does not satisfy all conditional expressions, then non-deterministic choices at steps 2 and 3 are unsuitable. 
\item For each $(q,a)$ such that $\sigma(\theta(q,a))=\emptyset$, construct $R_\sigma$ from $R$ by rewriting each $(q,a,e) \in R$ to $(q,a,\emptyset)$,
and check whether $TL(A_\sigma)$ is infinite where $A_\sigma=(Q,\Sigma,q_0,R_\sigma)$.
If $TL(A_\sigma)$ is infinite, then output ``Yes (i.e., $TL(Doc)$ is infinite).''
Otherwise, if there exists a pair $(q,a)$ such that $\sigma(\theta(q,a))$ is infinite
and there is $t = \langle T, l, \rho \rangle \in TL(A_\sigma)$ such that $(q,a) = (r^t_{A_\sigma}(v), l(v))$ for some $v \in T$,
then output ``Yes.''
If there does not exist such pair $(q,a)$, then output ``No (i.e., $TL(Doc)$ is finite).''
%Otherwise, if there exists $v \in T$
%such that for each $t = \langle T , l , \rho \rangle \in TL(A_\sigma)$ and its run $r_{A_\sigma}^t$
%$\sigma(\theta(r_{A_\sigma}^t(v),l(v)))$ is an infinite set,
%then output ``$TL(Doc)$ is infinite.''
%If there does not exist such $v \in T$, then non-deterministic choices in the steps 2 and 3 are unsuitable.
\end{enumerate}

\begin{figure}[htbp]
\centering
\rotatebox[origin=c]{0}{
\includegraphics[scale=.48]{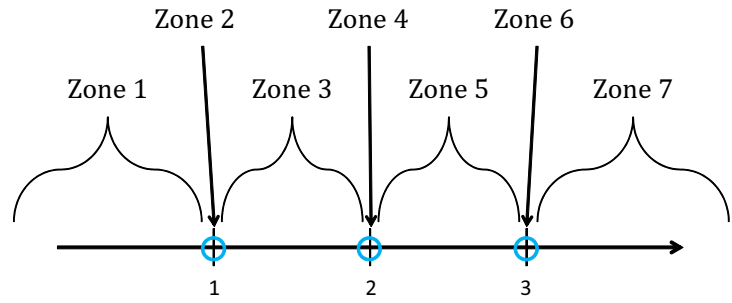}}
\caption{The breakup of the number line into zones.}
\label{fig:step1}
\end{figure}
\begin{figure}[htbp]
\centering
\rotatebox[origin=c]{0}{
\includegraphics[scale=.48]{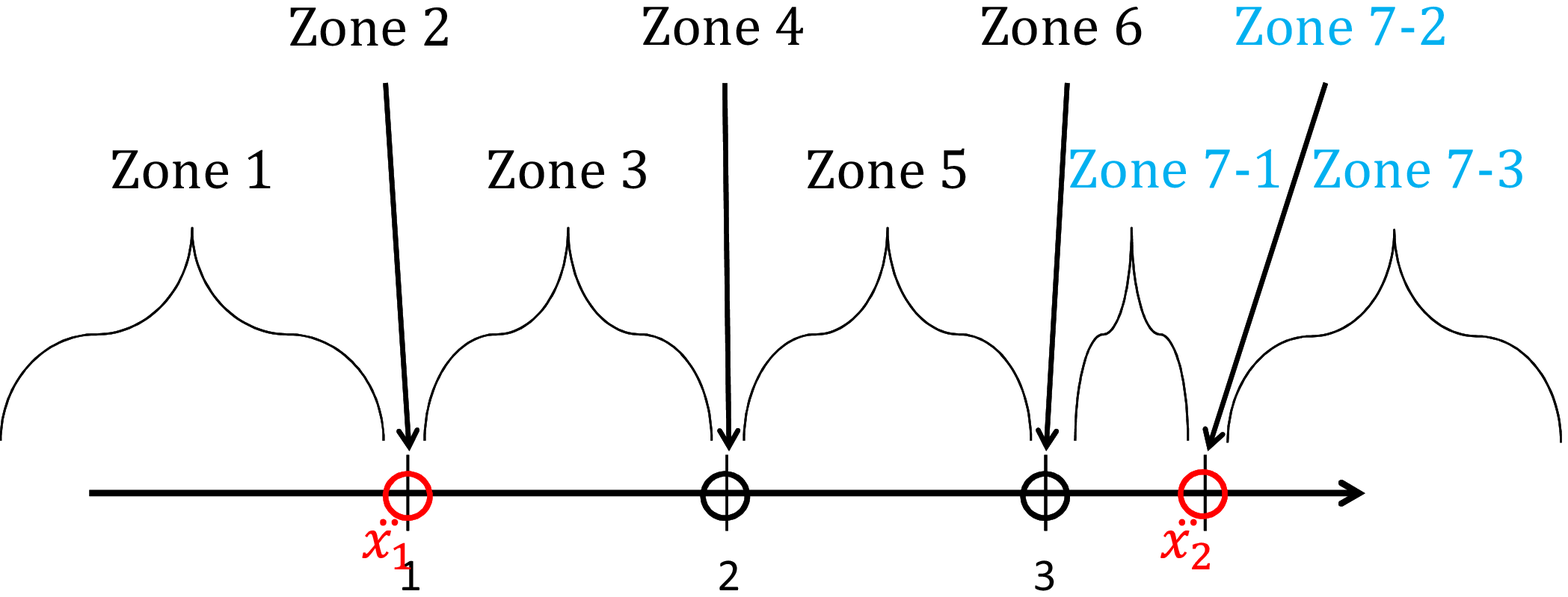}}
\caption{The assignment of zones to s-variable.}
\label{fig:step2}
\end{figure}
\begin{figure}[htbp]
\centering
\rotatebox[origin=c]{0}{
\includegraphics[scale=.48]{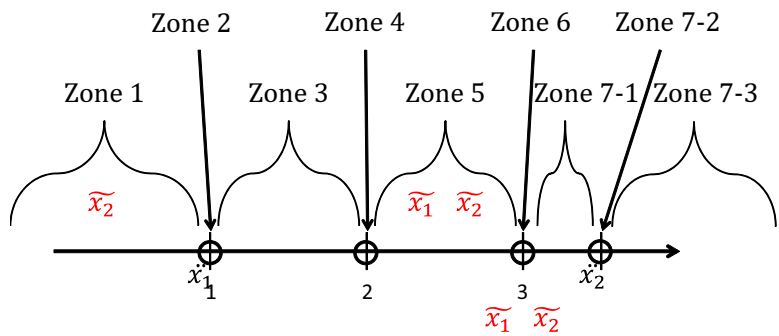}}
\caption{The assignment of zones to m-variable.}
\label{fig:step3}
\end{figure}

We show an example of an assignment and a breakup of zones.
Consider the set $E = \{ 1 \in \ddot{x_1}, 2 \not\in \ddot{x_2}, 3 \in \tilde{x_1}, \tilde{x_1} \subseteq \tilde{x_2} \}$.
First, as shown in Fig. \ref{fig:step1}, the number line is broken up into seven zones.
Next, as shown in Fig. \ref{fig:step2}, zone 2 and zone 7 are assigned to $\ddot{x_1}$ and $\ddot{x_2}$, respectively,
and then zone 7 is broken up into three zones 7-1, 7-2, and 7-3 by the assignment.
Finally, as shown in Fig. \ref{fig:step3}, zones 5 and 6 are assigned to $\tilde{x_1}$,
and zones 1, 5, and 6 are assigned to $\tilde{x_2}$.%decidability of infiniteness of data tree types

%ongoing and future work
\section{Ongoing and Future Work}
This paper has discussed security verification against inference attacks
on data trees. We have proposed tree transducers on data trees
which can represent projection,
selection, and natural join in the relational algebra.
Moreover, we have proposed data tree types for representing
the candidate set of the value of the sensitive information.

We are now trying to prove that type inference and inverse type inference are
possible on queries with respect to data tree types.
We have done inverse type inference on several tree transducers
and type inference on data-rewriting transducers until now.
One of our future work is to evaluate the complexity of our method.
Another future work is to consider inference attacks using functional
dependencies~\cite{HKIF12} on data trees.

%references
\bibliographystyle{eptcs}
\bibliography{main}
\end{document}